\documentclass[showpacs,preprint,amssymb]{revtex4}
\usepackage{amsfonts}
\usepackage{amssymb}
\usepackage{amsmath}
\usepackage{graphicx}
\usepackage{dcolumn}

\usepackage{bm}
\def\be{\begin{equation}} 
\def\ee{\end{equation}}
\def\bea{\begin{eqnarray}} 
\def\eea{\end{eqnarray}}
\def\line{\hbox to \hsize}    
\def\frac #1#2{{#1\over #2}}

\def\psid{\psi^{\dagger}}

\def\vev #1{{\langle #1\rangle}}
\def\1{\mbox{\bf 1}}
\def\bm#1{\mbox{\boldmath$#1$}} 

\begin{document}

\title{Berry phase and anomalous velocity of Weyl fermions  and Maxwell photons}

\author{ MICHAEL STONE}

\affiliation{University of Illinois, Department of Physics\\ 1110 W. Green St.\\
Urbana, IL 61801 USA\\E-mail: m-stone5@illinois.edu}   
\begin{abstract}  

We consider two  systems of wave equations whose  wave-packet solutions have trajectories that are altered by the ``anomalous velocity''  effect of a Berry curvature. The first is the matrix Weyl equation describing   cyclotron motion of a charged massless fermion. The second is  Maxwell equations for  the whispering-gallery modes of light in a cylindrical  waveguide.  In the case of the massless fermion,  the anomalous velocity is obscured by the contribution from the magnetic moment. In the whispering gallery modes  the anomalous velocity causes the  circumferential  light ray to creep up the cylinder at the rate of one wavelength per orbit, and can be identified  as a continuous version of the Imbert-Federov effect.
 \end{abstract}

\maketitle

\section{Introduction}

In many quantum  systems the motion of a wave-packet  is governed by   semiclassical equations of the form
\cite{blount,sundaram-niu,niu,horvathy}
 \bea
\dot {\bf k} &=&  -\frac{\partial {\mathcal H}}{\partial {\bf x}}+  e(\dot {\bf x}\times {\bf B}), 
\label{EQ:lorentz}\\
\dot {\bf x}&=&\frac{\partial {\mathcal H} } {\partial  {\bf k}}  + \dot {\bf k} \times {\bm \Omega}.
\label{EQ:chang-niu}
 \eea
 In the absence of the last term in the second equation  these would just be Hamilton's equations for a particle with hamiltonian ${\mathcal H}({\bf x},{\bf k})=\varepsilon({\bf k})+V({\bf x})$ moving in a magnetic field.  The  additional $\dot {\bf k}\times {\bm \Omega}$ term in (\ref{EQ:chang-niu})  is  the  {\it anomalous velocity\/} correction to the na{\"i}ve group velocity $\partial \varepsilon/\partial{\bf k}$.

The vector  ${\bf \Omega}$ is a function  of  the kinetic momentum ${\bf k}$ only, and is a   Berry curvature which  has  different origins in different systems. For  a  Bloch electron  in an energy band in a solid  the curvature   accounts   for  the effects of  all other  bands. In particle-physics applications  the  curvature arises from the  intrinsic angular momentum of the particle.  In all cases it   affects   the velocity  because  different momentum components of  a localized wave-packet accumulate different geometric phases when both  ${\bf k}$ is changing and the Berry curvature is non-zero   \cite{chong}. These ${\bf k}$-dependent  geometric phases are just as significant in determining the wave-packet position as  the ${\bf k}$-dependent dynamical phases  arising from  the dispersion equation $\omega=\varepsilon({\bf k})$. 
 
 A particularly simple example occurs in  in the dynamics of massless relativistic fermions and in Weyl semimetals where bands touch at a point.  In both systems  the wavepackets are solutions of a  Weyl ``half-Dirac'' equation and the Berry curvature arises because  the spin (or  pseudo-spin) vector is locked to the direction of the momentum ${\bf k}$. The forced precession of a spin-$S$ vector  is the paradigm in  the original Berry-phase paper \cite{berry1} and the corresponding curvature is simply
 \be
 {\bm \Omega} ({\bf k})= S \frac{{\bf k}}{|{\bf k}|^3}.
 \label{EQ:berry-curvature}
 \ee
Even  this basic example  gives rise to much physics --- the axial and gauge anomalies \cite{stephanov,stone-dwivedi,dwivedi-stone} and the chiral magnetic and vortical effect \cite{vilenkin,fukushima,son3}.
 
 The semiclassical analyses that reveal the anomalous velcity in  (\ref{EQ:chang-niu}) are quite intricate as they have to go beyond the leading  WKB ray tracing equation. It is the aim of this paper to consider  two   simple   systems   in which the predicted anomalous velocity effect can be sought directly in the stationary eigenfunctions of  underlying wave equation.  In both cases  the  curvature is given by (\ref{EQ:berry-curvature}). The first (in section \ref{SEC:cyclotron})  is the circular motion of a massless charged spin-1/2 particle in a magnetic field. The second (in section   \ref{SEC:whisper}) is the circular motion of a spin-1 photon in an optical fibre waveguide.  In the first case the presence of the anomalous velocity is obscured by the coupling of the  magnetic field to the particle's magnetic moment. The second unambiguously displays  the expected anomalous velocity drift.

 \section{Cyclotron orbits}
\label{SEC:cyclotron}

We start by considering  the cyclotron motion of  a massless Weyl fermion with positive charge $e$ in a magnetic field ${\bf B}= -B \hat{\bf z}$. The field  is derived from a   vector potential ${\bf A}= B(y,-x)/2$ and its  downward direction has been  chosen so that the particle orbits in an anti-clockwise direction about the $z$ axis. 

The Weyl Hamiltonian for a right-handed  spin-1/2 particle is 
\be
H= -i{\bm \sigma}\cdot (\nabla -ie{\bf A}),
\ee
where ${\bm \sigma}=(\sigma_1,\sigma_2,\sigma_3)$ denotes the Pauli matrices. We are using natural units in which $\hbar =c=1$, although we will occasionally insert these symbols when it helps to illuminate the discussion. 

 Acting on functions  proportional to $e^{ik_z z}$ we have  
\be
H^2 = {\mathbb I}\left(- \nabla^2 + \frac{e^2B^2}{4} r^2 +eBL_3 +k_z^2\right) + eB\sigma_3,
\label{EQ:weyl-squared}
\ee
where $L_z= -i(x\partial_y-y\partial x)=-i\partial_\phi$ is the canonical  (as opposed to kinetic) angular momentum and ${\mathbb I}$ denotes the 2-by-2 identity matrix.
The  eigenvalues of  the scalar Shr\"odinger operator  in parenthesis in (\ref{EQ:weyl-squared}) are  
\be
E^2_{n,l,k_z} = eB\left\{2n+|l|+l+ 1\right\}+k_z^2,
\ee
and the corresponding  eigenfunctions are
\be
\varphi_{n,l,k_z}(r,\phi) =\left(\frac{eB}{2}\right)^{(|l|+1)/2} \sqrt{\frac{n!}{(n+|l|)!}} r^{|l|}\exp\left(-\frac{eBr^2}{4}\right) L^{|l|}_n\left(\frac{eBr^2}{2}\right) e^{il\phi}e^{ik_z z}.
\ee
 Both $n$ and $l$ are integers and  $L^{|l|}_n$ is the associated Laguerre polynomial.
When $n=0$, $k_z=0$,  and $l>0$, the  wavefunction $\varphi_{0,l,0}(r,\phi)$  corresponds to a  particle describing a circular cyclotron orbit  with  the origin as its centre and radius 
\be
R_l=\sqrt{\frac{2l}{eB}}.
\ee
 If we  decrease $l$ while staying in the same Landau level ({\it i.e.\/}\  by increasing $n$ so  as to keep $E^2_{n,l}$ fixed)  the classical circular orbit  keeps the same radius but its centre moves away from the origin and is smeared-out in $\theta$ over the full  $2\pi$.  When $l=0$ the circle passes through the origin. For $l$ negative, the energy no longer depends on $l$ and the Landau level keeps $n$ fixed while  $l$ continues to decrease.  The classical  orbit still has the original radius, but no longer encloses  the origin.   
In particular  the case   $n=k_z=0$ and $l<0$    corresponds  particles in the lowest Landau level but with different orbit centres.

By  applying the   projection operator $P= (E+H)/2E$ to the Schr\"odinger eigenfunction we find that 
the  cyclotron-motion  eigenfunctions  of the Weyl hamiltonian $H$ with  $n=0$, $l>0$, and longitudinal momentum $k_z$ are  
\be
\psi_{0,l,k_z}(r,\theta,z)= e^{ikz}e^{il\phi}\left[\begin{matrix} (E_{l,k_z}+k_z)r^{-1} e^{-i\phi}\cr ieB  \end{matrix}\right] r^l \exp\left(-\frac{eB r^2}{4}\right).
 \ee
These states    have energy $E_{l, k_z} = \sqrt{2leB+k_z^2}$ and the  orbit radius is still
 \be
 R_l= \sqrt{\frac{2l}{eB}}.
 \ee
 At $k_z=0$, the angular velocity of a wave-packet is 
 \be
\dot \phi =   \left.\frac{\partial E_{l,k_z}}{\partial l}\right|_{k_z=0}=\sqrt{\frac{eB}{2l}}
 \ee
 ensuring that $v_\phi= R_l\dot \phi=c=1$.
 
 There is a special case where $l=n=0$ and 
 \be
 \psi_{0,0, k_z}=e^{ik_z z} \left(\begin{matrix}0\cr 1\end{matrix}\right) \exp\left(-\frac{eB |z|^2}{4}\right)
 \ee
 with $E=-k_z$. This mode only exists as a positive energy mode for $k_z<0$. It is this unbalanced mode, with a density of $eB/2\pi$ per unit area in the $x$, $y$ plane that is  the source of the chiral-magnetic-effect current 
  \be
  {\bf J}_{\rm CME}= \frac{e^2{\bf B}}{2\pi}\int_0^\mu \frac{dk_z}{2\pi}= \frac{e^2}{4\pi^2}{\bf B}\mu,
  \ee
  of a gas of zero-temperature Weyl fermions with chemical potential $\mu$ \cite{vilenkin,fukushima}.

Consider the $l>0$,  $k_z=0$ orbits.  Even though these orbits possess   no component of momentum in the $z$ direction,  plugging the time dependence of the classical orbital  momentum ${\bf k}$ into the  anomalous-velocity formula  (\ref {EQ:chang-niu}) suggests that  they should creep down  the $z$ axis.  
To compute the predicted creep-rate we observe that  a  particle of helicity $S$ whose spin direction is forced to describe a circle of co-latitude $\theta$ on a sphere with polar co-ordinates $\theta$, $\phi$ accumulates Berry phase at the rate \cite{berry1}
 \be
 \dot \gamma_{\rm Berry} =-S(1-\cos\theta)\dot \phi.
 \ee 
For $S=+1/2$, and using our expression for $\dot \phi$, this becomes 
\be
\dot \gamma_{\rm Berry}= \frac12(\cos \theta-1)\dot \phi = \frac12 (\cos \theta-1)\sqrt{\frac{eB}{2l}}.
\ee
where 
\be
\cos\theta = \frac {k_z}{\sqrt{2leB+k_z^2}}\sim \frac{k_z}{\sqrt{2leB}}.
\ee
In an energy eigenstate this  accumulating geometric phase should be indistinguishable from  the accumulating $-Et$ dynamical phase. In other words $\dot \gamma_{\rm Berry}$  should appear as   a contribution to the energy of 
\bea
 E_{\rm Berry}&=& {\rm const}. -  \frac 12 \frac {k_z}{\sqrt{2leB+k_z^2} }\sqrt{\frac{eB}{2l}}\nonumber\\
&\sim & {\rm const}. -\frac 12 \frac{k_z}{2l}.
\eea
This energy adds  
\be
\dot z= \frac{\partial E_{\rm Berry}}{\partial k_z}= -\frac 12 \frac{1}{2l}
\label{EQ:drift1}
\ee
to the group velocity over above that expected from the $E=c|{\bf k}|$  energy-momentum relation. 
The  velocity (\ref{EQ:drift1}) corresponds to  drift rate of  one-half of a de Broglie wavelength $\lambda_{\rm Broglie}$  per orbit --- the de Broglie wavelength being here defined as the  wavelength associated with the kinetic momentum ${\bf k}$ so that
$
E = {2\pi \hbar c }/{\lambda_{\rm Broglie}}
$.

Unfortunately, except for the special case  $l= 0$,  there is no sign of any contribution to the energy linear in $k_z$  in the exact solution of the eigenvalue problem!  Instead we have
\be
E=\sqrt{2leB+k_z^2}=  \sqrt{2leB}+ \frac 12 k_z^2/ \sqrt{2leB}+\cdots.
\ee
 The reason  for  the absence is that there  is  another linear-in-$k_z$ contribution to the energy coming from the Weyl particle's magnetic moment \cite{son3}.  

One might question whether  a massless particle can have  a magnetic moment as no time ever passes  on a null-vector world-line.  Nonetheless,  in the laboratory frame, a particle obeying the  Weyl   equation possesses an energy-dependent effective moment of 
$$
{\bm \mu}_{\rm Weyl}=\pm  \frac{e\hat{\bf k}}{2E}. \quad \hbox{($\pm $ for positive/negative  helicity)}.
$$
This moment  is precisely what is required for  the Larmor precession frequency 
$$
\Omega_{\rm Larmor}=  -B \left|\frac{\bm \mu}{\bf S}\right|
$$
to coincide with the orbital frequency and so  ensure that the   spin remains aligned with the momentum.  The appendix contains a   derivation of ${\bm \mu}_{\rm Weyl}$  from the Weyl equation  and explains how the Dirac gyromagnetic ratio $g=2$ continues to to be valid even for massless particles.
For a   positive-helicity particle the interaction of   this moment with the magnetic field   provides  an energy shift of  
$$
\delta E_{\rm magnetic}=-{\bm\mu}_{\rm Weyl}\cdot  {\bf B}= -\frac{e{\bf B}\cdot {\hat {\bf k}}}{2E}= \frac{eB\cos\theta}{2\sqrt{2leB}}=+\frac 12 \frac{k_z}{2l} .
$$
that  precisely cancels the Berry phase contribution.   This cancellation does not apply to all eigenstates and does not affect the total chiral magnetic effect, but   as explained in  \cite{son3}  it does complicate  its  simple semiclassical derivation  in \cite{stephanov}.

\section{Photons in  whispering-gallery modes}
\label{SEC:whisper}

We were   thwarted in our attempt to  observe  an  anomalous drift velocity in   an exact solution of the Weyl equation because the accumulating geometric  Berry phase  was   obscured  by a   dynamical phase  arising from  the particle's magnetic moment.  We   therefore   seek the effect in  the motion of a  moment-free massless spinning particle.  An  obvious candidate is the spin-1 photon.   To make it  clear that a photon should obey similar semiclassical equations to a  massless fermion  it helps to rewrite Maxwell's equations in a form that affords a  direct comparison with the Weyl equation. The idea for a such a rewriting is apparently due to Riemann. The complete  set of Maxwell's equations   equations  were first published in 1865 and Riemann died in  1866. The  rewriting  is nonetheless ascribed to him    by  Heinrich Weber in his edition  of  Riemann's lectures published in 1901 \cite{riemann}. The idea was independently discovered by  Silberstein \cite{silberstein} in 1907, and has been recently extensively championed by the Bia{\l}ynicki-Birula's \cite{birula}.  The Riemann-Silberstein equations make use the complex-valued fields 
\bea
{\bm \Psi}^{\pm}&=&  {\bf E}\pm i c{\bf B}\nonumber\\
&=& {\bf E}\pm i Z {\bf H},
\eea
where 
\be
Z=  \sqrt{\frac{\mu}{\epsilon}} 
\ee
is the wave impedance. Provided that $Z$ does not vary with position or time, we can 
combine  the two Maxwell ``curl" equations as
\be
i \partial_t  \Psi ^{\pm}_i = \pm  c_{\rm local}  \,\epsilon_{ijk} \partial_j \Psi ^\pm_k .   
\label{EQ:RS1}  
\ee
Here $c_{\rm local}\equiv 1/\sqrt{\mu\epsilon}$ may  vary with position.
At non-zero frequency, the two Maxwell ``divergence'' equations follow from curl equations and do not need to be separately imposed.  
Once we  define the spin-1 generator ${\bm \Sigma}_i$ to be  the matrix with entries $- i \epsilon_{ijk}$,  eq.~(\ref{EQ:RS1}) becomes a pair of   Weyl  equations
\be
i\hbar \partial_t {\bm \Psi}^\pm = \pm c_{\rm local}  \,({\bm \Sigma}\cdot \hat {\bf p}) {\bm \Psi}^\pm,
\ee
one for the left-helicity chiral field and one for the right-helicity  field. 
to make the analogy with the Dirac-Weyl equation  as close as possible, we  have inserted an $\hbar$ on both sides of (\ref{EQ:RS1}) so that we can exhibit the  equation  in terms of  the  quantum mechanical momentum operator $\hat {\bf p}= -i\hbar \nabla$.  
We conclude that as  the direction of the wave-momentum vector precesses, the photon spin is forced to follow it. The photon wavefunction will then  acquire a geometric Berry phase that is twice as a large as that of the Weyl fermion, and this phase must have a similarly-proportioned  effect on the semi-classical particle trajectory.

The  only significant difference  between the spin-$1/2$ Weyl  equation and   Maxwell equations  is  that  the  Maxwell  ``wave-function'' 
obeys a Majorana condition
\be
({\bm \Psi}^+)^*= {\bm \Psi}^-
\ee
that indicates  that the photon is its own antiparticle.

 \begin{figure}
\center{\includegraphics[width=3.0in]{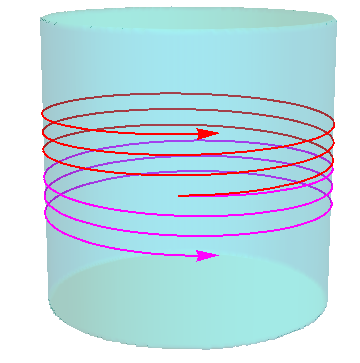}}
\caption{\sl The  separation of the positive helicity circumferential ray (magenta) from the negative helicity  ray (red)  in a cylindrical dielectric waveguide. Wavepackets drift up and down the fiber at the rate of one   wavelength per turn.}
\label{FIG:bliokh}
\end{figure}

To  obtain an optical  analogue of cyclotron motion, we  consider the  whispering-gallery modes of light  in   a step-index optical fibre.  
In whispering-gallery  modes a light beam  orbits the fibre circumferentially rather than propagating along its length.  We may  think of the orbit as a closely spaced sequence  of total internal reflections off the step discontinuity in the refractive index. (See figure \ref{FIG:bliokh}).  A separation of the  orbits of left and right circularly polarized beams  was predicted  from the anomalous velocity equations in    \cite{bliokh-bliokh} and experimentally verified   in  \cite{bliokh-verification}. In this section  we seek to derive the separation  directly from the  Maxwell-equation eigenmodes.

Recall that the refractive index $n$ is obtained from the material parameters as $n/c= \sqrt{\epsilon\mu}$ and, in natural units where $c=1$, the  local speed of light is $1/n$.  We take the  axis of the fibre  as   the $z$ axis and its  core to   have refractive index $n=n_1$ for $r<R$. The cladding  will have  index $n=n_2$ for $r>R$.    All  fields will  have a tacit factor of  $e^{ik_z z-i\omega t}$ so our   differential operators  act only on  functions of the transverse co-ordinates $x$, $y$.   We define parameters $\gamma$ by
\be
\gamma^2= \pm(\epsilon\mu \omega^2-k_z^2)
\ee
where the sign is chosen so that $\gamma$ is real.
 
The transverse  field components  ${\bf E}_\perp$ and ${\bf H}_\perp$ are expressed in terms of the longitudinal components   $E_z$ and $H_z$  by
 \bea
{\bf E}_\perp &=&\frac{i}{\gamma^2}\left\{k_z \nabla_\perp E_z - \omega \mu\, {\bf e}_z \times \nabla_\perp H_z\right\}, \nonumber\\
{\bf H}_\perp &=&\frac{i}{\gamma^2}\left\{k_z \nabla_\perp H_z + \omega \epsilon\,  {\bf e}_z \times \nabla_\perp E_z\right\}, 
\eea
where, for our step fibre (no gradients of $\epsilon$ or $\mu$  away from the discontinuity) 
\bea
-\nabla_\perp^2 H_z +(k_z^2 -\epsilon\mu \omega^2) H_z&=&0,\nonumber\\
-\nabla_\perp^2 E_z +(k_z^2 -\epsilon\mu \omega^2)E_z&=&0.
\label{EQ:wave-equations}
\eea
To be guided we need $k_z^2-\epsilon\mu \omega^2$ to be negative in the core, and positive  in the cladding. The solutions 
to (\ref{EQ:wave-equations}) are then
\be
E_z= 
\begin{cases} A{\rm J}_l(\gamma_1 r)e^{il\theta}
,&  r<R,\\
B {\rm K}_l(\gamma_2 r) e^{il\theta},  & r>R,
 \end{cases}
\ee
\be
H_z= 
\begin{cases} C{\rm J}_l(\gamma_i r)e^{il\theta}
,&  r<R,\\
D {\rm K}_l(\gamma_2 r) e^{il\theta},  & r>R.
 \end{cases}
 \ee
Here ${\rm J}_l$ and ${\rm K}_l$ are the Bessel function and modified Bessel function respectively. 

Below the fibre cutoff frequency the quantity $\gamma^2=\epsilon\mu \omega^2 -k_z^2$ is always positive.  Light is no longer completely confined, so the eigen-frequencies  have  a  negative imaginary that implies an exponential decay in time. The corresponding eigenfunctions  must have outgoing waves at infinity, and so will  be of the form
 \be
E_z= 
\begin{cases} A{\rm J}_l(\gamma_i r)e^{il\theta}
,&  r<R,\\
B {\rm H}^{(1)}_l(\gamma_2 r) e^{il\theta},  & r>R,
 \end{cases}
\ee
\be
H_z= 
\begin{cases} C{\rm J}_l(\gamma_i r)e^{il\theta}
,&  r<R,\\
D {\rm H}^{(1)}_l(\gamma_2 r) e^{il\theta}. & r>R,
 \end{cases}
 \ee
 Here ${\rm H}^{(1)}_l$ is a Hankel function of the first kind.

Whispering-gallery modes have  small $k_z$ and large azimuthal quantum number $l$. These modes are always below the fibre cutoff, and so $\gamma^2$ does not change sign at $r=R$. If we consider the waves near the point $(x,y,z)=(R,0,0)$ then we have 
\bea
E_x&=& \frac{i}{\gamma^2}\left( k_z \frac{\partial E_z}{\partial r}-\frac{\omega \mu }{r} \frac{\partial H_z}{\partial \theta}\right)\nonumber\\
E_y&=& \frac{i}{\gamma^2}\left( \frac{k_z}{r} \frac{\partial E_z}{\partial \theta}+{\omega \mu } \frac{\partial H_z}{\partial r}\right)\nonumber\\
H_x&=& \frac{i}{\gamma^2}\left( k_z \frac{\partial H_z}{\partial r}+\frac{\omega \epsilon }{r} \frac{\partial E_z}{\partial \theta}\right)\nonumber\\
H_y&=& \frac{i}{\gamma^2}\left( \frac{k_z}{r} \frac{\partial H_z}{\partial \theta}-{\omega \epsilon } \frac{\partial E_z}{\partial r}\right)
\eea
On substituting the functional form of the solutions  for $r<R$ these become
\bea
E_x&=& \frac{i}{\gamma_1^2}\left( k_z \gamma_1 {\rm J'}_l(\gamma_1r)A-\frac{il\omega \mu}{r} {\rm J}_l(\gamma_1 r) C\right)\nonumber\\
E_y&=& \frac{i}{\gamma_1^2}\left( \frac{ik_z l }{r} {\rm J}_l (\gamma_1 r)A +{\omega\mu }\gamma_1  {\rm J'}_l(\gamma_1 r) C\right)\nonumber\\
H_x&=& \frac{i}{\gamma_1^2}\left( k_z \gamma_1 J'(\gamma_1 r) C +\frac{i l \omega  \epsilon }{r} {\rm J}_l (\gamma_1 r)A \right)\nonumber\\
H_y&=& \frac{i}{\gamma_1^2}\left( \frac{i k_zl }{r} {\rm J}_l(\gamma_1 r) C -{\omega \epsilon  \gamma_1 } {\rm J'}_l(\gamma_1 r) A\right),
\eea
and for $r>R$
\bea
E_x&=& \frac{i}{\gamma_2^2}\left( k_z \gamma_2 {\rm H'}^{(1)}_l(\gamma_2r)B-\frac{il\omega\mu_2 }{r} {\rm H}^{(1)}_l(\gamma_2 r) D\right)\nonumber\\
E_y&=& \frac{i}{\gamma_2^2}\left( \frac{ik_z l }{r} {\rm H}^{(1)}_l(\gamma_2 r)B +{\omega\mu_2 }\gamma_2  {\rm H'}^{(1)}_l(\gamma_2r) D\right)\nonumber\\
H_x&=& \frac{i}{\gamma_2^2}\left( k_z \gamma_2  {\rm H'}^{(1)}_l(\gamma_2r) D +\frac{il \omega \epsilon_2  }{r} {\rm H}^{(1)}_l(\gamma_2 r)B \right)\nonumber\\
H_y&=& \frac{i}{\gamma_2^2}\left( \frac{i k_zl }{r} {\rm H}^{(1)}_l(\gamma_2 r)D -{\omega \gamma_2 \epsilon_2 }  {\rm H'}^{(1)}_l(\gamma_2r) B\right).
\eea
The boundary conditions are that the tangential components  $ E_z$, $E_y$  and $H_z$, $H_y$ be continuous. The continuity of the normal components of ${\bf D}$ and ${\bf B}$ is then ensured by the Maxwell ``curl'' equations. 

If $k_z =0$, then $\gamma= n \omega= \sqrt{\mu\epsilon} \,\omega $ and the boundary condition equations break into two blocks. The $E_z$ and $H_y$ continuity equations for one block are, respectively,  
\bea
 {\rm J}_l(n_1\omega R)A &=& {\rm H}^{(1)}_l(n_2 \omega R)B,\nonumber\\
\sqrt{\frac{\epsilon_1}{\mu_1}} J'(n_1\omega R)A &=& \sqrt{\frac{\epsilon_2}{\mu_2}}  {\rm H'}^{(1)}_l (n_2 \omega R) B. 
\eea
The $H_z$ and $E_y$ continuity equations for the other are 
\bea
 {\rm J}_l(n_1\omega R)C &=& {\rm H}^{(1)}_l(n_2 \omega R)D,\nonumber\\
\sqrt{\frac{\mu_1}{\epsilon_1}}J'(n_1\omega R)C &=& \sqrt{\frac{\mu_1}{\epsilon_2}}{\rm H'}^{(1)}_l (n_2 \omega R) D.
\eea
There are therefore  two  families of whispering-gallery   modes. The first is comprises  TM modes (transverse when looking along  the fibre) that have $H_z= 0$ (and therefore non-zero $A$, $B$) with frequencies determined by the eigenvalue equation
\be
\sqrt{\frac{\mu_1}{\epsilon_1}}  \frac{{\rm J}_l(n_1 \omega R)}{{\rm J'}_l(n_1 \omega R)}=  \sqrt{\frac{\mu_2}{\epsilon_2}} \frac{{\rm H}^{(1)}_l(n_2 \omega R)}{{\rm H'}^{(1)}_l (n_2 \omega R)}.
\label{EQ:TM}
\ee
 The second comprises   TE modes with $E_z=0$ (and therefore non-zero $C$, $D$), with
 \be
\sqrt{\frac{\epsilon_1}{\mu_1}}  \frac{{\rm J}_l(n_1 \omega R)}{{\rm J'}_l(n_1 \omega R)}= \sqrt{\frac{\epsilon_2}{\mu_2}}  \frac{{\rm H}^{(1)}_l(n_2 \omega R)}{{\rm H'}^{(1)}_l (n_2 \omega R)}.
\label{EQ:TE}
\ee
 Equivalently the eigenfrequencies are  the zeros of    
\bea
{\rm Det_{TM}}(\omega R) &\equiv & \sqrt{\epsilon_1/\mu_1} {\rm J'}_l(n_1 \omega R){\rm H}_l^{(1)}(n_2 \omega R) -  \sqrt{\epsilon_2/\mu_2}   {\rm H'}^{(1)}(n_2 \omega R){\rm J}_l(n_1 \omega R)\nonumber\\
{\rm Det_{TE}}(\omega R) &\equiv & \sqrt {\mu_1/\epsilon_1}{\rm J'}_l(n_1 \omega R){\rm H}_l^{(1)}(n_2 \omega R) - \sqrt{\mu_2/\epsilon_2} {\rm H'}^{(1)}(n_2 \omega R){\rm J}_l(n_1 \omega R).
\eea
In general  TE and TM  modes with the same $l$  are not degenerate.

Consider first the case of $n_1<n_2$.  This is not the situation  in a practical  optical fibre where the core always has a higher refractive index, but there are still low-$Q$ resonant modes whose partial  confinement arises  because grazing-angle incidence provides strong reflection. In this  case, as $r$ increases, the Hankel functions  are approaching  their asymptotic oscillating region 
\be
{\rm H}^{(1)}_l(n_2 \omega r ) \sim \sqrt{\frac{2}{\pi n_2 \omega r}} e^{in_2 \omega r }+\ldots 
\ee
before  ${\rm J}_l(n_1 \omega  r)$ starts to oscillate. For  $l$ large and $n_1/n_2 \approx 1$ the RHS of equations (\ref{EQ:TM}) and (\ref{EQ:TE}) are slowly varying functions of $x=n_1\omega r$. They  have a very small real part and a negative imaginary part.  Consequently the eigenfrequencies of both TE and TM modes are close to zeros of ${\rm J}_l(x)$ but we must   give $x$ a  negative imaginary part to make the determinants vanish.   The eigenfrequencies are therefore given by 
$$
n_1\omega R= z_{l,n}= \xi_{n,l}- i \eta_{l,n}
$$
where $\xi_{l,n}$ is the $n$-th zero of ${\rm J}_l(x)$ and $\eta_{l,n}$ is a positive quantity  that differs for TE and TM modes.

\begin{figure}
\center{\includegraphics[width=4.0in]{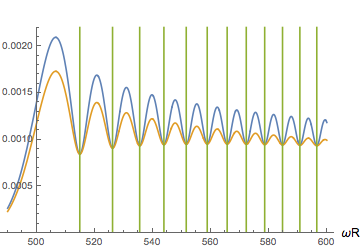}}
\caption{\sl Plot of  the magnitudes of ${\rm Det_{TM}}(\omega R)$ (blue curve) and  ${\rm Det_{TE}}(\omega R)$ (brown curve) as a function of $\omega R$ for $l=500$. The refractive indices are  $n_1=1.0$, $n_2=1.1$. The green lines are the locations of the first few zeros of  ${\rm J}_l(n_1 \omega R)$. The locations of the minima are the real parts of $R$ times the  resonant eigenfrequencies.  }
\label{FIG:unconfined}
\end{figure}

Now  we introduce a small $k_z$. The pairs of equations are now coupled, but  we see that the term with $k_z$ on the LHS of the $E_y$ and $H_y$  continuity equations   multiplies ${\rm J}_l(x)$, and this quantity is small at resonance.  We therefore neglect these terms. If we look at the  $E_x$ and $H_x$ continuity  equations (whose validity was previously enforced by the others) and neglect the RHS radiation fields, we find that the  eigenvalue equations simplify to 
\bea 
k_z n_1 \omega {\rm J'}_l(n_1\omega R) A -\frac{il \omega} {R} {\rm J}_l(n_1\omega R)C&=&0,\nonumber\\
 k_z n_1 \omega {\rm J'}_l(n_1 \omega R) C+\frac{il \omega n^2}{R} {\rm J}_l(n_1\omega R)A&=&0.
\label{EQ:reduced-eqs-1}
 \eea
As  we are very close to zeros of the Bessel function, we can set  ${\rm J}_l(z_{l,n}+n_1\delta \omega R)\approx  n_1 R \,{\rm J'}_l(z_{l,n})\,\delta \omega$
 and (\ref{EQ:reduced-eqs-1})  reduce to
 \bea
 k_z A- il \,\delta \omega\, C&=&0,\nonumber\\
 k_z  C +il  n_1^2 \,\delta\omega\, A&=&0. 
 \label{EQ:reduced-eqs-2}
 \eea
 These two equations coincide once  we set $(A,C)=(1,\pm i n_1)$, which   corresponds to right and left circularly polarized light in a medium of refractive index $n_1$. 
They give a  frequency shift of   
 \be
 \delta \omega = \pm \frac{1}{l n_1} k_z. 
 \ee
 The longitudinal group velocity of a wave packet centered around $k_z=0$ s therefore 
 \be
 \dot z \equiv \frac{\partial \omega}{\partial k_z}= \pm \frac{1}{ n_1 l}.
 \ee
 Since $n_ 1\omega R\sim l$ for these modes and $n_1 \omega = k_\phi$ is the  wavenumber for the light in the core, we can write this equation as 
 \be
 \dot z = \pm \frac{1}{n_1 k_\phi R}= \pm \frac{ \lambda_{\rm glass} }{2\pi n_1 R} 
 \ee
 where $\lambda_{\rm glass}$ is the wavelength of the light in the fibre core. 
 As the time for one orbit is   $2\pi R n_1$, we find that  the rate of drift is 
  one (in glass) wavelength per orbit. 
This  is exactly what we expect from the berry phase argument in section \ref{SEC:cyclotron}.  The photons  can only make a few orbits, however,  before they escape  or become depolarized due to the  TE mode (which, looking along the circumferential ray, is a linearly polarized beam with the ${\bf E}$ field in the radial direction)   having a shorter lifetime that the TM (which is a linearly polarized beam with the ${\bf E}$ field in the $z$ direction).

 In an actual fibre   we have $n_1>n_2$. In this regime   the Bessel function ${\rm J}_l(n_1\omega r)$ begins to oscillate while the Hankel function 
 ${\rm H}^{(1)}_l(n_2 \omega r )$ is still almost real  and exponentially  decreasing. In the region of  $\omega R$ corresponding to  total internal reflection,  the  fields outside the glass  decay almost to zero as they would for total internal reflection  off a flat  interface ---  but eventually the Hankel function  begins  to oscillate and the fields become outgoing radiation.

 We assume that $k_z$ is small, so that we can ignore the $k_z$'s in  $\gamma_1$ and $\gamma_2$.    The boundary condition equations are then 
 \bea
 \left( \frac{ik_z l}{R} \frac{1}{\epsilon_1 \mu_1} \,{\rm J}_l(\gamma_1R ) C - \omega^2\sqrt{\frac{\epsilon_1}{\mu_1}} {\rm J'}_l(\gamma_1 R) \,A \right)&=&
 \left( \frac{ik_z l}{R} \frac{1}{\epsilon_2 \mu_2} \,{\rm H}_l^{(1)}(\gamma_2R ) D - \omega^2\sqrt{\frac{\epsilon_2}{\mu_2}} {{\rm H}'}_l^{(1)}(\gamma_2 R) \,B \right)\nonumber\\
{\rm J}_l(\gamma_1R)\,A &=& {\rm H}^{(1)}_l(\gamma_2 R )\,B \nonumber\\
 \left( \frac{ik_z l}{R} \frac{1}{\epsilon_1 \mu_1} \,{\rm J}_l(\gamma_1R ) A - \omega^2\sqrt{\frac{{\mu}_1}\epsilon_1} {\rm J'}_l(\gamma_1 R) \,C \right)&=&
 \left( \frac{ik_z l}{R} \frac{1}{\epsilon_2 \mu_2} \,{\rm H}_l^{(1)}(\gamma_2R ) B - \omega^2\sqrt{\frac{\mu_2}{\epsilon_2}} {{\rm H}'}_l^{(1)}(\gamma_2 R) \,D \right).\nonumber\\
  {\rm J}_l(\gamma_1R)\,C &=& {\rm H}^{(1)}_l(\gamma_2 R )\,D
 \eea 
  
\begin{figure}
\center{\includegraphics[width=4.0in]{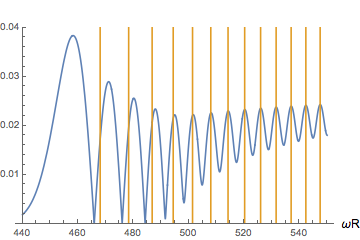} }
\caption{\sl Plot of   the magnitude  of ${\rm Det}(\omega R)/{\rm H}_l^{(1)}(n_2 \omega R)  $  (blue curve)  against $\omega R$ for an  impedance-matched fibre with $n_1=1.1$, $n_2=1$ and $l=500$. The brown  lines are the locations of the first few zeros of  ${\rm J}_l(n_1 \omega R)$. The three deep minima  are resonances in the  high-$Q$ total-internal-reflection region.  The minima  become shallower  once the angle of incidence of the rays decreases below  the critical angle, allowing the  light to  leak out and  the eigen-frequencies to acquire a significant  imaginary part. }
\label{FIG:symmetric}
\end{figure}

 From  now on we make the  assumption  that the impedance  does not change from core to cladding--- {\it I.e.\/}\ that  $\sqrt{{\mu_1}/{\epsilon_1}}=  \sqrt{{\mu_2}/{\epsilon_2}}$. This impedence matching condition might  be hard to engineer, but  given that we are seeking a mathematical illustration  of the anomalous-velocity equation rather  than proposing experimental verification   it is not unreasonable. The   matching  means that the right and left handed  Riemann Silberstein fields do not mix. It also  ensures  the   degeneracy of the  $k_z=0$ TE and TM modes,
 both eigenfrequences being    determined  by the same  vanishing  condition 
  $$
 {\rm Det}(\omega R) \equiv  {\rm J'}_l(n_1 \omega R){\rm H}_l^{(1)}(n_2 \omega R) -   {\rm H'}^{(1)}(n_2 \omega R){\rm J}_l(n_1 \omega R)=0.
 $$
 See figure \ref{FIG:symmetric} for a plot that locates the zeros of ${\rm Det}(\omega R)$.
 
 \begin{figure}
\center{\includegraphics[width=4.0in]{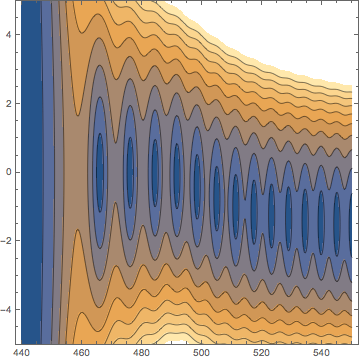}}
\caption{\sl Contour plot   (blue is lowest value) of  the magnitude  of ${\rm Det}(\omega R)/{\rm H}_l^{(1)}(n_2 \omega R)  $  In the complex $\omega R$ plane for the same range of parameters as figure \ref{FIG:symmetric}. The first three zeros lie very close to the real axis.  They correspond to modes that are strongly confined by total internal reflection. The remaining zeros break away from the axis and their larger  imaginary parts  indicate   that the light  rays can escape because they are below the critical angle.}
\label{FIG:contour}
\end{figure}

For non-zero $k_z$ the TE and TM pairs of equations are again coupled, but, accepting  the impedance-matching condition  we can decouple the four equations into a different two  pairs of equations --- one for each helicity. We set 
 \bea
 (A,C)&=& (1, i \sqrt{\epsilon_1/\mu_1})X_+\nonumber\\
 (B,D)&=&(1,i \sqrt{\epsilon_2/\mu_2}) Y_+
 \eea
 and similarly, with the sign before the $i$ changed,  for  $X_-$,  $Y_-$.
 Using the notation $c_{1,2}=1/n_{1,2}$, the equations for the ``+'' pair  become 
 \bea
 \left(\frac{k_zl}{R} c_1^2 {\rm J}_l(\omega R/c_1) +\omega^2 {\rm J'}_l(\omega R/c_1)\right) X_+ &=& \left(\frac{k_zl}{R}{c_2}^2 {\rm H}_l^{(1)}(\omega R/c_2) +\omega^2 
 {{\rm H}'}_l^{(1)}(\omega R/c_2)\right)Y_+\nonumber\\
{\rm J}_l (\omega R/c_1)X_+&=& {\rm H}_l^{(1)}(\omega R/c_2)Y_+
 \eea
 Dividing the first by the second equation and rearranging gives 
 \be
 \frac{k_z l}{R}(c_1^2-{c_2}^2) =\omega^2\left(\frac{ {\rm J'}_l(\omega R/c_1)}{{\rm J}_l(\omega R/c_1)}  - \frac{{{\rm H}'}_l^{(1)}(\omega R/c_2)}{{\rm H}_l^{(1)}(\omega R/c_2)}\right).
 \ee
 When $k_z=0$ the vanishing of the RHS is the eigenvalue condition. To find $\partial\omega/\partial k_z$ at $k_z=0$, we therefore  need to compute  the derivative of the RHS at the  points at which it vanishes.  To do this we make use of   the asymptotic formula
 
\begin{figure}
\center{\includegraphics[width=4.0in]{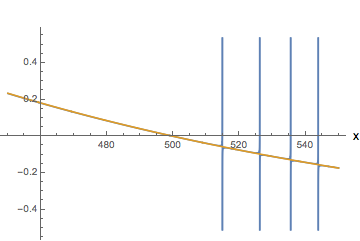}}
\caption{\sl Plot of ${\rm J}''_l(x)/{\rm J}_l(x)$  (blue curve) compared with  $l^2/x^2-1$ (brown curve) for $l=500$. The brown curve sits on top of the blue curve except for the spikes near  the zeros of ${\rm J}_l(x)$.}
\label{FIG:bessel}
\end{figure}
  \begin{figure}
\center{\includegraphics[width=4.0in]{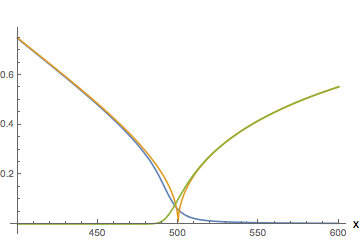}}
\caption{\sl Plot of the real (blue curve) and imaginary parts (green curve) of $\,{\rm H'}^{(1)}_l(x)/{\rm H}^{(1)}_l(x)$  compared with   $\sqrt{|1-l^2/x^2|}$ (brown curve) for $l=500$.}
\label{FIG:hankel}
\end{figure}

 \be
 \frac {{\rm J}''_l(x)}{{\rm J}_l(x)}\sim \frac{l^2}{x^2}-1\nonumber\\
 \ee
 which is accurate for large  $x$ provided that we stay away from places where ${\rm J}_l(x)$ vanishes [see figure \ref{FIG:bessel}]. This condition is satisfied at the points of interest [see figure \ref{FIG:symmetric}]. We may also use  the formul\ae\
 \bea
 \frac{{{\rm H}''}_l^{(1)}(x)}{  {\rm H}_l^{(1)}(x)}&\sim& \frac{l^2}{x^2}-1\nonumber\\
 \frac{{{\rm H}'}_l^{(1)}(x)}{  {\rm H}_l^{(1)}(x)}  &\sim& - \sqrt{\frac{l^2}{x^2}-1},\quad x<l.
 \eea
 These last two approximations [see figure \ref{FIG:hankel}] imply  that
 \be
 \left(\frac{{\rm H}'}{\rm H}\right)' = \frac{{\rm H}''}{\rm H}-\left(\frac{{\rm H}'}{\rm H}\right)^2\sim  0,
 \ee
 which is not quite right, but the derivative on the LHS  is  $O(1/l)$ in the region of interest and can be neglected.  We also note that we can evaluate
 \be
  \left(\frac{{\rm J}'_l(x)}{{\rm J}_l(x)}\right)' = \frac{{\rm J}_l''(x)}{{\rm J}_l(x)}-\left(\frac{{\rm J}_l'(x)}{{\rm J}_l(x)}\right)^2
 \ee
 at the unperturbed eigenvalue by exploiting the fact that  ${\rm J}_l'/{\rm J}_l={\rm H}'_l/{\rm H}_l$ at that point. 
 We find 
 \bea
  \frac{l}{R}(c_1^2-{c_2}^2) \delta k_z &=& \frac{\omega^2 R}{c_1} \left[\left(\frac{l^2c_1^2}{\omega^2 R^2}-1\right)- \left(\frac{l^2 c_2^2}{\omega^2 R^2}-1\right)\right]\delta \omega \nonumber\\ 
  &=& \frac{\omega^2 R}{c_1} \left[ \frac{l^2(c_1^2-{c_2}^2)}{\omega^2 R^2} \right]\delta \omega.
  \eea
  Thus  
  \be
  \delta \omega = \frac{c_1}{l} \delta k_z= \frac{c}{n_1 l}\, \delta k_z.
  \ee
  The opposite circularly-polarized have an opposite frequency shift. We have therefore recovered the same drift equation
  \be
 \dot z= \frac{\partial \omega}{\partial k_z}= \pm \frac{c}{n_1 l}
  \ee
   that we found for the $n_1<n_2$ fibre. 
Recall that this  drift is at a rate of one wavelength per orbit.

\section{Discussion} 

We can compare the  mode-expansion rate of drift  with that expected from angular momentum conservation about an axis perpendicular to the reflection interface.  For a particle of momentum ${\bf p}$ and  helicity $S$, and treating   the orbit as a series of grazing-angle reflections,    each deflection through an  angle  $\delta \phi $ causes a  change in the perpendicular spin component of $2S \sin(\delta \phi/2)$ that must be compensated for by a change in the orbital angular momentum of $-|{\bf p}| \delta z$. Thus $|{\bf p}| \dot z= - S\dot \phi$.   Since $\dot\phi R= c/n_{\rm glass}$ we have  
 \be
 \dot z = - S \frac{c}{n_{\rm glass} |{\bf p}|R}.
 \ee
 In this picture the drift may  be  understood as a continuous version of the Imbert-Fedorov effect \cite{imbert,federov}.
 
 Agreement  with the results of the previous section  requires us to identify the magnitude of the photon momentum $|{\bf p}|$ with $2\pi \hbar/\lambda_{\rm glass} =n_{\rm glass} \hbar \omega/c$, This is the Minkowski expression for the momentum of a photon in a  medium --- as   
 opposed to  Abraham's  expression for the momentum  which places   the $n_{\rm glass}$ in the denominator. (For a review the Abraham-Minkowski momentum controversy see \cite{abraham-minkowski}.)
 Minkowski's   momentum is today   understood to be the {\it pseudo-momentum\/}  which is conserved as a result of the homogeneity of the medium \cite{blount-pseudomomentum}. It is the rotational symmetry of the medium about the normal to the core-cladding interface that is responsible for the angular momentum conservation, so the appearance of the Minkowski momentum is not surprising.  
 
 That a continuous process of reflection and rotation can transport a circularly polarized beam of light  parallel to itself through an arbitrary distance is  related  to the fact that a continuous sequence of Lorentz boosts and rotations in free space can  perform an arbitrary   ``Wigner translation" of a finite  polarized beam \cite{stone-wigner}.

 \section{Acknowledgements}   This  project was supported by the National Science Foundation  under grant    NSF DMR 13-06011.   I would  like to thank  Konstantin Bliokh  and Misha Stephanov for  e-mail discussions.
 
 \section{Appendix:  Gordon decomposition, the Weyl magnetic moment,  and $g=2$.} 
  The  orginal  Gordon decomposition of the Dirac 4-current  \cite{gordon}  shows  that for any  solution $\psi$  of the  massive Dirac equation 
 \be
 (i\gamma^\mu (\nabla_\mu-m)\psi=0,
 \label{EQ:Dirac-massive}
 \ee
 the four-current can be expressed as
 \be
 \bar\psi \gamma^\mu\psi =\frac{i}{2m} (\bar \psi \nabla^\mu\psi -(\nabla^\mu\bar \psi) \psi)+\frac{1}{m} \partial_\nu(\bar\psi  \Sigma^{\mu\nu}\psi),
 \ee
 where 
 \be
 \Sigma^{\mu\nu} = \frac {i}{4} [\gamma^\mu,\gamma^\nu]
 \ee
 is the Lorentz generator. 
 
Gordon's  decomposition of the current into a particle number-flux and bound spin contribution  clearly  requires $m\ne 0$. There is, however,  a version that is valid in both massive and massless cases: 
 assume that  $\psi({\bf r},t)=\psi({\bf r})\exp\{-iEt\}$  and make use of the Dirac equation in the Hamiltonian form 
 \bea
 \partial_t\psi &=& -{\bm \alpha}\cdot \nabla\psi -im\beta \psi, \nonumber\\
 \partial_t \bar\psi&=& +\nabla\bar\psi\cdot {\bm \alpha} +i m\bar\psi \beta,\nonumber
 \eea
 with $\beta=\gamma^0$,  $\alpha^i=\gamma^0\gamma^i$ and so find  
 \be
{\bf j}\equiv   e\bar \psi {\bm \gamma} \psi = \frac{e}{2iE} \left(\psid \nabla  \psi - (\nabla \psid)\psi\right) +\frac{e}{E} (\nabla \times{\bf  S}).
 \ee
 Here ${\bm \gamma}= (\gamma^1,\gamma^2,\gamma^3)$, and 
 \be
 {\bf S} =\psid \hat {\bf S}\psi
 \ee
with 
\be
(\hat S_x,\hat S_y,\hat S_z)= (\Sigma^{23},\Sigma^{31},\Sigma^{12})
\ee
 so that 
 \be
 \hat {\bf S}=\frac 12 \left[\begin{matrix}{\bm \sigma}&0 \cr 0 &{\bm \sigma}\end{matrix}\right].
 \ee
 With the particle-number density identified with $\rho= \psid\psi$, we can again interpret the first term in the decomposition as the  current ${\bf j}_{\rm free}= e\rho {\bf k}/E= e\rho {\bf v}$ due to particles moving at speed ${\bf v}={\bf k}/E$. The second term, ${\bf j}_{\rm bound}= (e/E)\nabla\times {\bf S}$ is the current due to the gradients in the intrinsic magnetic moment density.  The  magnetic  moment itself is found by integrating by parts to show that  
 \be
 {\bm \mu}\stackrel{\rm }{=} \frac{1}{2}\int {\bf r}\times {\bf j}_{\rm bound}\,d^3x  =\frac{1}{2}\int {\bf r}\times \left(\frac e E\nabla \times {\bf S}\right)\,d^3 x = \frac{e}{E}\int {\bf S}\,d^3 x. 
 \ee
 For a single  massless particle whose spin-1/2  is locked to the direction  $\hat {\bf k}$ of its  kinetic momentum this is 
 \be
 {\bm \mu}_{\rm Weyl}= \frac{e \hat {\bf k}}{2E},
 \ee
as claimed in section \ref{SEC:cyclotron}.
 
For the both massive and massless case we also have  an expression for the momentum density as part of the symmetric  Belinfante-Rosenfeld energy-momentum tensor 
 \be
 T^{\mu\nu}_{\rm BR}= \frac{i}{4}(\bar \psi \gamma^\mu \nabla^\nu \psi - (\nabla^\nu \bar\psi) \gamma^\mu\psi +\bar \psi \gamma^\nu \nabla^\mu \psi-(\nabla^\mu \bar\psi) \gamma^\nu\psi).
 \ee
 Using the Dirac equation we evaluate $T^{0\mu}_{\rm BR}=({\mathcal E},{\bf P})$ to find  ${\mathcal E}=E\psid \psi$, and 
 \be
 {\bf P}= \frac 1{2i}\left (\psid (\nabla \psi)- (\nabla \psid)\psi\right) +\frac 12 \nabla\times {\bf S}.
 \ee
(If we used the non-symmetric canonical energy-momentum tensor 
 \be
 T^{\mu\nu}_{\rm canonical}= \frac{i}{2}(\bar \psi \gamma^\mu \nabla^\nu \psi - (\nabla^\nu \bar\psi) \gamma^\mu\psi),
 \ee
 we do would not  find  the bound spin-momentum contribution.)
 
 Again integrating by parts, we  recover the spin contribution to the total angular momentum density as 
 \be
  \int {\bf r}\times\left(\frac 12 \nabla\times {\bf S}\right)\,d^3x = \int {\bf S}\, d^3x,
 \ee
 so the division by 2 in the spin contribution  to the momentum density is correct. The  absence of a division by 2 in the formula for the current  reflects the $g=2$ gyromagnetic ratio of the electron. In other words a  spin-density  gradient is twice as effective at making an electric current as it is at contributing to the  momentum.

\end{document}